\documentclass[aps,prb,twocolumn,psfig,showpacs]{revtex4}
\usepackage{amsfonts}
\usepackage{amsmath}
\usepackage{graphicx}
\usepackage{bm}
\usepackage{amssymb}
\usepackage{times}
\usepackage{dcolumn}
\usepackage{cases}
\usepackage{txfonts}
\usepackage{color}

\newcommand {\Y}{\textcolor {blue}}

\RequirePackage[hyperindex,colorlinks,bookmarksnumbered,plainpages=true]{hyperref}
\hypersetup{colorlinks,linkcolor=blue,urlcolor=blue,citecolor=blue}
\usepackage{hyperref}

\usepackage{ulem}
\renewcommand{\emph}[1]{\textit{#1}}

\definecolor{darkgreen}{rgb}{0,0.5,0}
\definecolor{darkblue}{rgb}{0,0,0.5}
\definecolor{darkred}{rgb}{.7,0,0}
\definecolor{purple}{rgb}{0.35,0,0.35}
\definecolor{orange}{rgb}{1,0.5,0}
\definecolor{grey}{rgb}{.6,.6,.6}

\newcommand{\Eqs}[1]{Eqs.~(\ref{#1})}

\begin{document}

\title{Efficient simulation of infinite tree tensor network states on the Bethe lattice}
\author{Wei Li$^{1}$, Jan von Delft$^{1}$, and Tao Xiang$^{2,3}$}

\affiliation{$^1$Physics Department, Arnold Sommerfeld Center for Theoretical Physics, and Center for NanoScience, Ludwig-Maximilians-Universit\"at, 80333 Munich, Germany}

\affiliation{$^2$Institute of Physics, Chinese Academy of Sciences, P.O. Box 603, Beijing 100190, China}

\affiliation
{$^3$Institute of Theoretical Physics, Chinese Academy of Sciences, P.O. Box 2735, Beijing 100190, China}

\begin{abstract}
  We show that the simple update approach proposed by Jiang
  et al [H.C. Jiang, Z.Y. Weng, and T. Xiang,
  Phys. Rev. Lett. \textbf{101}, 090603 (2008)] is an efficient and
  accurate method for determining the infinite tree tensor network
  states on the Bethe lattice. Ground state properties of the quantum
  transverse Ising model and the Heisenberg XXZ model on the Bethe
  lattice are studied. The transverse Ising model is found to undergo
  a second-order quantum phase transition with a diverging magnetic
  susceptibility but a finite correlation length which is
  upper-bounded by $1/\ln(q-1)$ even at the transition point ($q$ is
  the coordinate number of the Bethe lattice). An intuitive
  explanation on this peculiar ``critical'' phenomenon is given. The
  XXZ model on the Bethe lattice undergoes a first-order quantum phase
  transition at the isotropic point. Furthermore, the simple update
  scheme is found to be related with the Bethe approximation. Finally, by
  applying the simple update to various tree tensor clusters, we can
  obtain rather nice and scalable approximations for two-dimensional
  lattices.
\end{abstract}

\maketitle

\section{introduction}

The investigation of quantum lattice models is one of the
central themes in modern condensed matter physics. A crucial
  step is to develop novel numerical methods to
efficiently simulate the interesting and complex phenomena of quantum
many-body systems. In particular, the tensor network states and the
related renormalization group methods, including the tree tensor
network state (TTN),\cite{Shi, Tagliacozzo, Murg} the multi-scale
entanglement renormalization ansatz,\cite{MERA} the projected
entangled pair state,\cite{Verstraete} the tensor renormalization
group (TRG),\cite{Levin, Xiang, Gu} and the second renormalization
group (SRG),\cite{Xie, Zhao} are now under rapid development. These
methods provide promising numerical tools for studying strongly
correlated systems, especially for the frustrated magnetic systems and
fermion models, and can be regarded as an extension of the fruitful
density matrix renormalization group (DMRG)\cite{White} in two or
higher dimensions.

\begin{figure}[tbp]
\includegraphics[angle=0,width=1.0\linewidth]{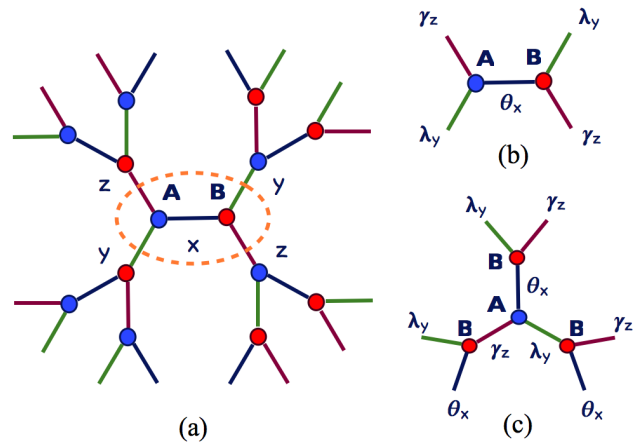}
\caption{(Color online) (a) The $q=3$ Bethe lattice. Every site has 3 nearest neighbors, and the three bonds are labeled according to their directions as $x$, $y$, and $z$, respectively.  (b) The two-site cluster used in the single-bond projection of the simple update scheme. The diagonal matrices $\lambda_y$ and $\gamma_z$ on the dangling bonds should be included in the projection to mimic the entanglement renormalization of the environment to this two-site system. (c) A minimum cluster that is used in the Bethe approximation. It consists of one A tensor and three nearest-neighbor B tensors (or vice versa).
} \label{fig-bethe-latt}
\end{figure}

In the study of the tensor network methods, one needs to first
determine the tensor network wavefunction for the ground state.  In
Refs. [\onlinecite{Xiang, Zhao}], a simple update scheme is proposed
to determine the ground state tensor network wavefunction in two
dimensions. This scheme is efficient and robust.  It proceeds in three steps:
(1) apply the imaginary time projection operators
simultaneously on bonds of the same type, for
example the $x$ directional bonds in Fig. \ref{fig-bethe-latt}a,
and enlarge the bond dimension; (2) construct a local evolving
block matrix and simulate the environment contribution by the diagonal
matrices on the external bonds [$\lambda_y$ and $\gamma_z$ in
Fig. \ref{fig-bethe-latt}b]; (3) decompose the evolving block
matrix by singular value decomposition (SVD) and decimate the vector
space of the enlarged geometric bond according to the singular values
in the updated diagonal matrix $\theta'_x$.  This technique has been
combined with the TRG/SRG to evaluate the ground state properties of
two-dimensional (2D) Heisenberg models.\cite{Zhao, Gu, Chen, Li,
  Zhao2} It is an accurate numerical method for evaluating local
physical quantities, but it is less accurate in evaluating the
long-range correlation functions.\cite{Zhao} This is the major
drawback of this simple update scheme.  It results from a mean-field
approximation for the environment tensor.  A way to go beyond this
approximation is to enlarge the size of the cluster that is used for
evaluating the environment tensor. This, as shown by Wang and
Verstraete\cite{Wang}, can indeed improve the accuracy for the long
range correlation function.

In this work, we apply the simple update scheme to infinite
TTN (iTTN) states on the Bethe lattice. We will show that
this is a quasi-canonical approach for treating an iTTN. Here by the
word ``quasi-canonical'' we mean that with increasing the number of iteration
steps and decreasing the Trotter error, the tree tensor network state
obtained by the simple update scheme would become asymptotically
canonical [i.e.\ the tensors satisfy certain canonical
  orthonormality conditions, see \Eqs{eq-can3} below]. Thus the
simple update scheme provides an accurate and efficient approach for
evaluating the ground state wavefunction on the Bethe lattice.

The Bethe lattice, as shown in Fig. \ref{fig-bethe-latt}a, has a
self-similar structure with an infinite Hausdorff dimension. The size
of the lattice is infinite, hence the boundary effects do not need to be explicitly
considered. The Bethe lattice was first used in the study of classical
statistical mechanics.\cite{Bethe, Plischke, Baxter} It has attracted
broader interest since a number of chemical compounds with the Bethe
lattice structures, such as the dendrimers,\cite{Delgado} have been
synthesized in the laboratory.\cite{Astruc}

A finite Bethe lattice is called a Cayley tree.  Soon after White's 
invention of DMRG,\cite{White} the DMRG algorithm
for the quantum lattice models defined on the Cayley tree was
proposed. \cite{Otsuka, Friedman} Based on the DMRG calculation of
local physical quantities in the central part of the Cayley tree,
Otsuka\cite{Otsuka} claimed that the anisotropic S=1/2 Heisenberg
model (i.e. the XXZ model) on the Bethe lattice should exhibit a
first-order phase transition at the isotropic point. Later
Friedman\cite{Friedman} proposed an improved DMRG scheme and evaluated
the spin-spin correlations in the ground state. Based on the DMRG
result, he suggested that long-range magnetic order might exist at
the isotropic Heisenberg point. Recently, Kumar et al. calculated the
magnetization with a further improved DMRG algorithm, \cite{Kumar} and
showed that such long-range magnetic order does exist at that point.

The above DMRG calculations were done on the Cayley tree lattice, not
on the true infinite Bethe lattice.  Furthermore, it should be pointed out
that the boundary effect is very strong on a finite Cayley tree since
more than one-half of the total sites reside on the lattice edge. This
may strongly affect the properties of the system. In some cases, the
results obtained on a Cayley tree lattice can be completely different from 
those for the corresponding Bethe lattice. For example,
the classical Ising model shows a phase transition on the Bethe
lattice, but not on the Cayley tree lattice.\cite{Ostilli}

To unambiguously resolve the above problems, it
is necessary to calculate the spin models directly on the Bethe
lattice.  The recent development of the TTN algorithms \cite{Shi,
  Tagliacozzo, Murg} has indeed made this feasible.\cite{Nagaj, Nagy}
In particular, Nagaj \textit{et al} in Ref. \onlinecite{Nagaj}
extended the infinite time evolving-block decimation\cite{Vidal}
technique to the Bethe lattice and determined the ground state
wavefunction by imaginary time evolution. For the transverse Ising
model on the Bethe lattice, it was found that a second-order quantum
phase transition exists at a critical transverse field. An interesting
result revealed in this calculation is that even at the second-order
critical point, the correlation length remains finite. For the Bethe
lattice with coordination $q=3$, the correlation length is shown to be
less than $1/\ln2$.  However, in the calculation Nagaj \textit{et al}
used a three-site projection operator to simultaneously evolve the two
equivalent incoming legs of the tensors, the computational cost is
thus very high. The computational time scales as O($D^8$) with $D$ the
tensor dimension, which limits the value of $D$ that can be handled to
$D\le 8$.

Recently, Nagy\cite{Nagy} proposed a different algorithm to reduce the computational cost by making use of the $C_3$ rotational symmetry of $q=3$ Bethe lattice. This algorithm reduces the computational cost to O($D^4$) hence greatly improves the efficiency. It can be used for studying the spin-1/2 quantum lattice models. However, the application of this method is restricted to the translation invariant spin-1/2 system.

As will be shown below, the simple update scheme is very efficient. Its computational costs scale as O($D^4$), similar as for the algorithm proposed by Nagy\cite{Nagy}. But it is much more flexible. It can be applied to treat arbitrary TTN states, with or without translation invariance. Here we studied two spin models defined on the Bethe lattice. One is the transverse Ising model and the other is the antiferromagnetic XXZ Heisenberg model. The quantum phase transitions and the ground state phase diagrams of these models are studied.

The rest of the paper is arranged as follows. An introduction to the
simple update scheme and its relationship with the Bethe approximation
is presented in Sec. \ref{sec:2}. The study of the quantum phase
transitions of transverse Ising and XXZ Heisenberg models are presented
in Secs. III and IV, respectively. In Sec. V, the present scheme is
generalized to larger tree tensor clusters, in order to provide more
accurate approximations for 2D lattices. Finally, Sec VI is devoted to a summary.

\section{The canonical form and the simple update scheme}
\label{sec:2}

The iTTN state on the Bethe lattice comprises 4-indexed tensors $A_{x,y,z}^m$ and $B_{x,y,z}^m$ defined on the vertices, and the diagonal matrices $\theta$, $\lambda$, $\gamma$ defined on the bonds along the $x$, $y$, and $z$ directions as shown in Fig. \ref{fig-bethe-latt}a, respectively.
The bond indices represent the quantum numbers of the virtual basis states. The physical index $m$ runs over the $d$ basis states of the local Hilbert space at each lattice site. The diagonal matrices store the entanglement information, and play an important role in the simple update scheme.

In order to determine the ground state wavefunction, the imaginary time evolving operators $U(\tau) = \exp{(-\tau h_{i,j})}$ are applied to the iTTN iteratively.
At each step, the dimension of the evolved bond is increased by a factor of $d^2$.
Thus the tensor dimensions will proliferate exponentially with the increasing number of projection steps.
In order to sustain the projections until the iTTN converges to the true ground state wavefunction, one needs to truncate the bond dimension after each projection step. This needs a proper consideration of the renormalization effect of the environment tensor.

An accurate and full determination of the environment tensor is
computationally costly.  This limits generally the tensor dimension
$D$ that can be handled to a rather small value, say $D \le 6$.  The
simple update scheme\cite{Xiang, Xie, Zhao}, on the other hand, takes
the product of the dangling bond matrices as a mean field
approximation to the environment tensor.  It converts the complicated
global optimization problem into a local one, and
hence greatly simplifies the calculation.  On the
regular 2D lattice, the bond matrix is an approximate measure of the
entanglement between the system and environment tensors.  However, as
will be shown later, the square of the diagonal bond matrix on the
Bethe lattice is the eigenvalue of the reduced density matrix if the
iTTN is canonicalized, i.e., if the tensors in the network are
always kept in canonical form by some transformations. 
Thus the simple update scheme is an accurate treatment for
the renormalization of the iTTN on the Bethe
lattice.

The simple update scheme is also closely related to the famous Bethe
approximation.\cite{Bethe, Plischke, Baxter} To understand this, we
show in Fig. \ref{fig-bethe-latt}c a 4-site cluster, which contains
one A tensor and three B tensors.  In the simple update calculation,
the two local tensors, A and B, and the three inner bond matrices
($\theta_x$, $\lambda_y$, $\gamma_z$) should be evaluated and updated
iteratively. After each single projection step on the inner bonds, to
keep the scheme self-consistent, one should also update all the
dangling bonds of the cluster, by replacing the bond matrices with the
corresponding ones on the inner bonds.

This cluster structure and the self-consistent scheme is in fact the
Bethe approximation that was first proposed by Bethe in 1930's, in the
context of statistical mechanics.\cite{Bethe} The key idea is to treat
the correlations between the central spin and its nearest neighbors in
the cluster exactly, and to use an effective mean field to approximate
the interactions between the cluster and the rest lattice spins.  By
solving this simple cluster problem, and assuming that all the spins
in the cluster have exactly the same local magnetization, one can
determine the spontaneous magnetization self-consistently.  For the
quantum cases, the six diagonal matrices on the dangling bonds of the
cluster are taken as the mean fields acting on the inner block. The
self-consistent condition requires that the matrices $\theta$,
$\lambda$, and $\gamma$ on dangling bonds are equal to the corresponding
matrices on the inner bonds between A and B tensors (see
Fig. \ref{fig-bethe-latt}c).

\begin{figure}[tbp]
\includegraphics[angle=0,width=1.0\linewidth]{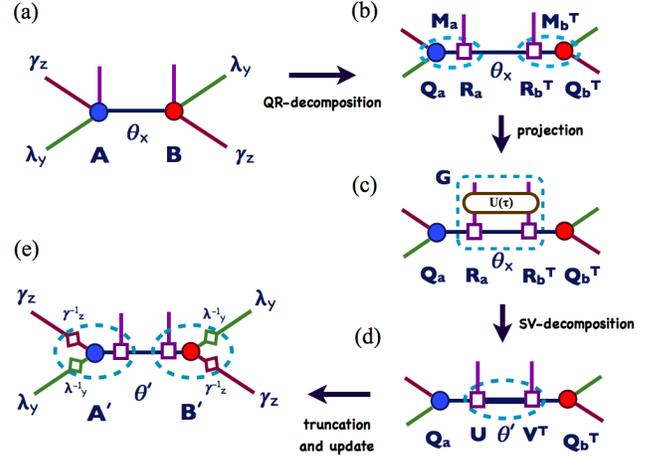}
\caption{(Color online) One iteration step in the simple update scheme:
(a) A and B tensors are connected by the bond $x$, which will be involved in the following projection steps.
There are diagonal matrices on the dangling bonds ($y$ and $z$ bonds) of A and B tensors.
(b) Absorb the four dangling matrices into A and B, and define the block matrix $M_{a(b)}$. 
Then take the QR decomposition for $M_{a(b)}$, obtaining $Q_{a(b)}$ and $R_{a(b)}$ matrices.
(c) Project $U(\tau)$ onto the bond by contractions, and obtain the block matrix $G$ [see Eq. (\ref{eq-G})].
(d) Take singular value decomposition of $G$ to find the unitary matrices $U$ and $V^{T}$,
and the new diagonal matrix $\theta'$.
(e) Truncate the $x$-bond dimension to $D$ according to the diagonal values of $\theta'$.
Merge $U$ ($V^T$), $\gamma_y^{-1}$, and $\lambda_z^{-1}$ together into $Q_{a(b)}$,
finally we arrive at the updated $A'(B')$ tensors.}
\label{fig-CTN}
\end{figure}

A tensor network state contains redundant gauge degrees of freedom on
each bond.  It is invariant if one inserts a product of two reciprocal
matrices on a bond and absorbs separately each of them to a local
tensor at the two ends of the bond.  This gauge invariance of a tensor
network state can be used to simplify the calculation of local
tensors, especially for the iTTN states on the Bethe lattice, where a
special gauge, called canonical form, can be introduced.

To be specific, the local tensors of canonical iTTN states satisfy the
following orthonormality conditions 
\begin{eqnarray}
\sum_m 
\sum_{x,y} \theta_x^2 \lambda_y^2 (T^{m}_{x,y,z'})^* T^{m}_{x,y,z} 
&=& \delta_{z',z},
\label{eq-can1}
\\
\sum_m
\sum_{y,z} \lambda_y^2 \gamma_z^2 (T^{m}_{x',y,z})^* T^{m}_{x,y,z} 
&=& \delta_{x',x}, \\
\sum_m
\sum_{z,x} \gamma_z^2 \theta_x^2 (T^{m}_{x,y',z})^* T^{m}_{x,y,z} 
&=& \delta_{y',y},
\label{eq-can3}
\end{eqnarray}
where T represents the A or B tensor. If we cut an arbitrary bond to
divide the Bethe lattice into two parts, denoted as a system and an
environment subblock, one can then define the reduced density matrix
of the system block by integrating out all the degrees of freedom
of the environment block. For the tensors that satisfy
Eqs. (\ref{eq-can1}-\ref{eq-can3}), the square of the diagonal bond matrices are the eigenvalues, and
the renormalized bond bases are the eigenvectors of the corresponding
reduced density matrix, which are orthonormal to each other. Thus, in
terms of the Schmidt decomposition, the square of the diagonal matrix
elements represent the probability amplitudes of the corresponding
eigenvectors appearing in the wavefunction.

The existence of this simple canonical form of the iTTN,
i.e. Eqs. (\ref{eq-can1}-\ref{eq-can3}), is very useful in the
calculations.  First, the diagonal bond matrix describes the
entanglement spectrum between the system and environment subblocks.
Thus to select the virtual bond basis states according to the values
of these diagonal matrix elements provides an optimal scheme to
truncate the bond dimension.  Second, the contribution of the
environment tensors can be faithfully represented by the 4 diagonal
matrices on the dangling bonds surrounding the central bond under
projection (see Fig. \ref{fig-CTN}a).  It means that the imaginary
time evolution on each bond can be done rigorously and locally.
Furthermore, we can also evaluate the expectation value of a local
operator simply by contracting a small cluster comprising those
tensors and bond matrices on which the operator acts. 
This significantly reduces the computational cost.

Bearing in mind the benefits of the canonical iTTN states, one
can perform explicitly the canonical transformations during the projection
processes. However, to further save computational costs, 
in practical calculations we choose to carry out the
imaginary time evolution just using the simple update scheme,
and gradually reduce the Trotter step $\tau$, which would bring
the iTTN states into its canonical form step by step.
This scheme works because the diagonal bond matrix provides an
approximate measure for the entanglement between the two sides of the
bond and can be used to substitute approximately the environment
tensor.  Therefore, it can stabilize the algorithm of the imaginary time
evolution, provided that the Trotter step $\tau$ is small enough so
that the bond projection operator $U(r)$ is nearly
unitary.\cite{Vidal} This near unitary evolution can
modify the wavefunction and reshape it in order to satisfy the canonical conditions.
The simple update scheme hence provides a quasi-canonical
evolution of the iTTN state, which will finally converge to the ground state
and become canonical in the limit $\tau\rightarrow 0$. In practical calculations, 
the Trotter step $\tau$ is gradually decreased from $10^{-1}$ to $10^{-4}$, and the total number
of projections steps varies from 20000 to 200000.

Now let us consider how to implement the simple update scheme
efficiently.  A simple approach is to do directly the singular value
decomposition the evolving block tensor, which is a matrix of $D^2
\times D^2$.  The computational cost for doing this singular value
decomposition is high, since it scales as
$O(D^6)$. This cost can in fact be reduced to $O(D^4)$ if we carry
out this singular value decomposition in the following steps (again,
projection on $x$ bond is taken as an example):

(1) Define the following two $D^2 \times D d$ block
matrices (Fig. \ref{fig-CTN}b)
\begin{eqnarray}
(M_a)_{y,z; x,m} & = & \lambda_y \gamma_z A_{xyz}^m, \\
(M_b)_{y,z; x,m} & = & \lambda_y \gamma_z B_{xyz}^m,
\end{eqnarray}
by absorbing the diagonal matrices $\lambda_y$ and $\gamma_z$ into the tensors A and B, and calculate their QR decomposition
\begin{equation}
(M_\alpha)_{y,z; x, m} =  \sum_{k} Q^\alpha_{y,z; k} R^\alpha_{k;m,x},
\end{equation}
where $\alpha = a$ or $b$. $Q^\alpha$ is a $D^2 \times Dd$ column orthonormal matrix. $R^\alpha$ is a $Dd \times D d$ upper diagonal matrix.

(2) Apply the bond projection operator $U(\tau)$ to
the system and define the gate matrix (Fig. \ref{fig-CTN}c)
\begin{equation}
G_{m_1k_1;m_2k_2} = \sum_{x, m_1^\prime, m^\prime_2}
\langle m_1 m_2 | U(\tau ) |m_1^\prime m^\prime_2\rangle
R^a_{k_1;m_1^\prime x} \theta_x R^b_{k_2;m_2^\prime x}
\label{eq-G}
\end{equation}
(3) Take the singular value decomposition for this matrix (Fig. \ref{fig-CTN}d),
\begin{equation}
G_{m_1k_1;m_2k_2} = U_{m_1k_1;l} \theta^\prime_l V_{m_2k_2; l},
\end{equation}
where $U$ and $V$ are two $Dd\times Dd$ unitary matrices, and
$\theta^\prime$ is a semi-positive defined matrix.

(4) Truncate the inner bond
dimension by keeping the largest $D$ matrix elements of
$\theta^\prime$, and update the local tensors by
the formula (Fig. \ref{fig-CTN}e)
\begin{eqnarray}
{A^\prime}^m_{xyz} &=& \sum_k  \lambda_y^{-1} \gamma_z^{-1} Q^a_{y,z;k} U_{m,k;x}, \\
{B^\prime}^m_{xyz} &=& \sum_k  \lambda_y^{-1} \gamma_z^{-1} Q^b_{y,z;k} V_{m,k;x}.
\end{eqnarray}

Combining this efficient simple update scheme and the local
determination of physical observables using the canonical form, we can
keep the computational cost in a low level. In practice, this
allows us to keep a relative large bond dimension. 

\begin{figure}[tbp]
\includegraphics[angle=0, width=0.85\linewidth]{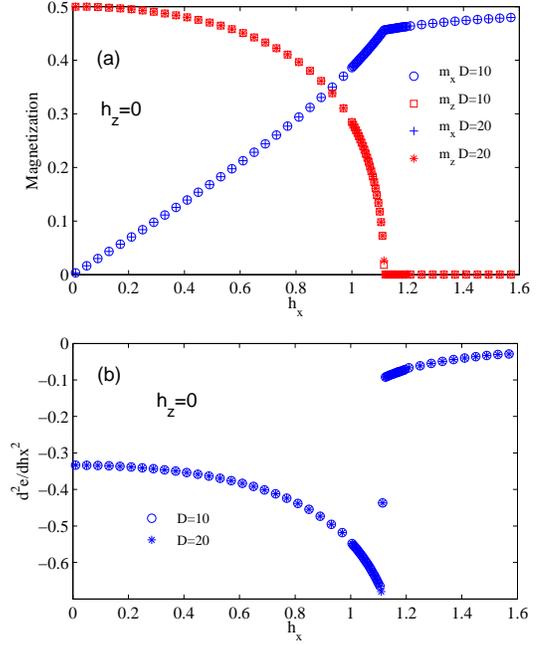}
\caption{(Color online)  (a) The longitudinal and transverse magnetizations $m_z$ and $m_x$ versus the transverse fields $h_x$. The transverse magnetization $m_x$ increases monotonously with $h_x$, while the longitudinal magnetization $m_z$ decreases and vanishes at the transition point. (b) The second-order derivative of the ground state energy per site $e$ with respect to $h_x$, $d^2 e/ dh_x^2$, obtained by taking the first-order derivative of $m_x$, $dm_x/dh_x$.}
\label{fig-magn-q-Ising}
\end{figure}

\section{The transverse Ising model}

The transverse Ising model is defined by the Hamiltonian
\begin{equation}
H_{\rm{TI}} = - \sum_{<i,j>} J S_i^z S_j^z - \sum_{i} h_x S_i^x - \sum_{i} h_z S_i^z,
\end{equation}
where the spin-spin exchange constant $J$ is set as the energy scale ($J=1$, ferromagnetic coupling).
The second term represents the transverse-field along $S^x$ direction, and the last term is the longitudinal-field along $S^z$ direction.

\begin{figure}[tbp]
\includegraphics[angle=0, width=0.85\linewidth]{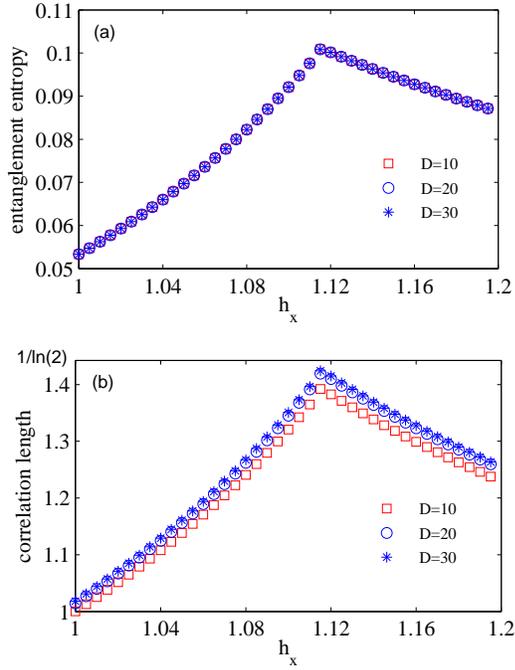}
\caption{(Color online) (a) The entanglement entropy $S_{\rm{E}}$ of the ground state for the transverse Ising model. The cusp at $h_c \simeq 1.115$ corresponds to the second-order phase transition point. (b) The correlation length $\xi$ of the ground state, which also shows a cusp at the transition point.}
\label{fig-ent-q-Ising}
\end{figure}

Figure \ref{fig-magn-q-Ising}a shows the longitudinal and
transverse magnetizations, $m_z = \langle S_z \rangle$ and $m_x =
\langle S_x \rangle$, as a function of the transverse field $h_x$. A
continuous order-disorder phase transition is found at $h_c$. For $h_x
< h_c$, the ground states undergo a spontaneous $Z_2$ symmetry
breaking with a finite longitudinal magnetization $m_z$, which
decreases with increasing $h_x$ and vanishes at the critical field. By
utilizing the Hellmann-Feynman theorem, the second-order derivative of
the ground state energy can be calculated by $d^2e/dh_x^2 =
-dm_x/dh_x$. As shown in Fig. \ref{fig-magn-q-Ising}b, $d^2e/dh_x^2$
exhibits a discontinuity at $h_c$, indicating that $h_c$ is a
second-order phase transition point.  The critical field is found to
be $h_c \simeq 1.115$, in agreement with previous
calculations.\cite{Nagaj, Nagy} It is also close to the critical field
$h_c = 1.06625(2)$ for the transverse Ising model on the honeycomb
lattice.\cite{Deng}

\begin{figure}[tbp]
\includegraphics[angle=0, width=0.9\linewidth]{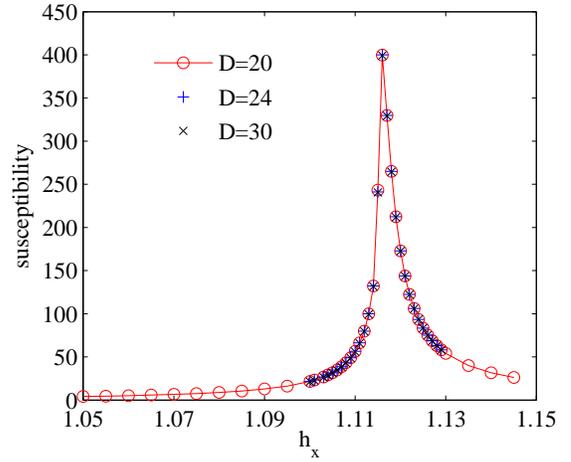}
\caption{(Color online) The field dependence of the longitudinal susceptibility $\chi^z$. It shows a divergent peak at the transition point $h_c \simeq 1.116$. The susceptibility is calculated by $\chi^z = [m_z(h_z)-m(0)]/\delta h$ with $h_z = 10^{-4}$.} \label{fig-sus-q-Ising}
\end{figure}

Figure \ref{fig-ent-q-Ising} shows the bipartite entanglement entropy $S_{\rm{E}}$ \begin{equation}
S_{\rm{E}} = - \rm{Tr}[\Lambda^2 \log_2(\Lambda^2)],
\label{eq:entropy}
\end{equation}and the correlation length $\xi$ for the ground state. In Eq. (\ref{eq:entropy}), $\Lambda $ = $\theta$, $\lambda$ or $\gamma$ is a diagonal matrix that satisfies the canonical condition. $S_{\rm{E}}$ shown in Fig. \ref{fig-ent-q-Ising}a is obtained by taking the average over the three bonds.

The correlation length $\xi$ is evaluated from the ratio of the the largest ($a_0$) and the second largest eigenvalue ($a_1$) of the transfer-matrix for the iTTN state,
\begin{equation}
\xi = 1/\ln{\frac{a_0}{a_1}} \; .
\end{equation}
For the $q=3$ Bethe lattice, there are six kinds of transfer-matrices, depending on
the site and the bond direction. For example the transfer matrix along
the $yz$-direction is defined by (for A sublattice site)
\begin{equation}
T^{a}_{y,z; y',z'}  =  \sum_{m,x} \sqrt{\lambda_y \gamma_z} (A^m_{x,y,z})^* \theta^2_x A^m_{x,y',z'} \sqrt{\lambda_{y'} \gamma_{z'}}.
\end{equation}
The other five transfer matrices including $T^b_{y,z;y',z'}$, $T^{a,b}_{z,x; z',x'}$, and $T^{a,b}_{x,y; x',y'}$ can be similarly defined. The results of the correlation length shown in Fig. \ref{fig-ent-q-Ising}b are evaluated from the product of the six transfer matrices along a specific path.

A distinctive feature revealed by Fig. \ref{fig-ent-q-Ising} is that
the correlation length $\xi$, as well as the entanglement
$S_{\rm{E}}$, does not diverge at the critical point. $\xi$ is found
to be upper bounded by $1/\ln{2}$, in agreement with the published
results\cite{Nagaj, Nagy}. This peculiar behavior is not observed in
the ordinary continuous phase transition systems, where the
correlation length is always divergent at the critical point.

We will now show that this non-critical behavior of the correlation
length at the critical point is due to the peculiar geometry of the Bethe lattice.
For the Bethe lattice, the number of sites on the boundary of
a finite connected region is roughly equal to the number of
internal sites within that region. It means that the lattice sites are highly
non-uniformly distributed as a function of lattice distance away from
a given center. This is a feature of the Bethe lattice that differs
from a regular lattice.

In order to understand why the correlation length is finite at the
critical point, let us take a scaling transformation to convert the
Bethe lattice to a ``regular'' 2D lattice whose lattice sites are
uniformly distributed in space. To do this, we first choose an arbitrary site,
to be viewed as ``center'' of the lattice, and define the distance $R$ for a given layer $r$ to the
center as
\begin{equation}
R \propto \sqrt{ \frac{N(r) }{ \pi } } \sim \sqrt{ \frac{(q-1)^r }{ \pi} }
\end{equation}
where $N(r) \propto (q-1)^r$ is the number of sites enclosed by the
$r$ layer.  In this rescaled lattice, $ r \sim 2\ln{R} + \mathrm{const}
$, and the exponentially decaying correlation
function $C(r) \sim \exp(-r/\xi)$ in the original Bethe lattice
corresponds to a power-law decaying function of $R$
\begin{equation}
C( R ) \sim R^{-2/\xi}.
\end{equation}
The algebraic decay of this correlation function
suggests the spins on the Bethe lattice are actually long range
correlated in the rescaled framework, even away from the critical
point. This is the reason why the system can undergo a phase
transition without exhibiting a divergent correlation length
at the critical point in the original Bethe lattice.

To understand why the correlation length is upper bounded, let us
consider the longitudinal magnetic susceptibility $\chi_z$ of the ground state (Fig. \ref{fig-sus-q-Ising})
\begin{equation}
\chi_z = \frac{d m_z}{d h_z} = \sum_{i} \langle S_0^z S_i^z \rangle_{\rm{GS}} - \langle S_0^z \rangle_{\rm{GS}} \langle S_i^z \rangle_{\rm{GS}},
\end{equation}
where $S_0^z$ is the spin at a reference center, $i$ runs over all the sites on the lattice. 
$\langle \hat{O} \rangle_{\rm{GS}}$ is the expectation value of operator
$\hat{O}$ in the ground state. Exploiting the $C_3$ rotational symmetry
of the Bethe lattice, we define $C^z(r) = \langle S_0^z S_i^z \rangle_{\rm{GS}} - \langle S_0^z
\rangle_{\rm{GS}} \langle S_i^z \rangle_{\rm{GS}}$, with $r$ the layer number
where site $i$ resides. Thus $\chi_z$ can be rewritten as
\begin{equation}
\chi_z = \sum_r n(r) C^z(r),
\label{eq-chiz}
\end{equation}
where $n(r)$ is the number of spins on layer $r$.  On a regular
lattice, $n(r) \propto r^{\nu-1}$, where $\nu$ is the spatial
dimension of the lattice, the susceptibility is always finite if the
spin-spin correlation function $C^z(r)$ decays exponentially.
However, in the Bethe lattice, $n(r) \propto (q-1)^{r-1}$. Now if we
assume $C^{z}(r) \propto \exp(-r/\xi )$, then
\begin{equation}
\chi_z \propto \sum_r (q-1)^r e^{-r / \xi} = \sum_r e^{r \left[\ln (q-1) - 1/\xi\right]},
\end{equation}
which diverges if $\xi$ approaches the threshold value $1/\ln(q-1)$. This shows that the
susceptibility can diverge even if $C^{z}(r)$ decays exponentially
with $r$. The critical point occurs when $\xi = 1/\ln (q-1)$,
and the correlation length $\xi$ is therefore upper-bounded by
$1/\ln (q-1)$ on the Bethe lattice. \cite{Semerjian}

\begin{figure}[tbp]
\includegraphics[angle=0, width=0.85\linewidth]{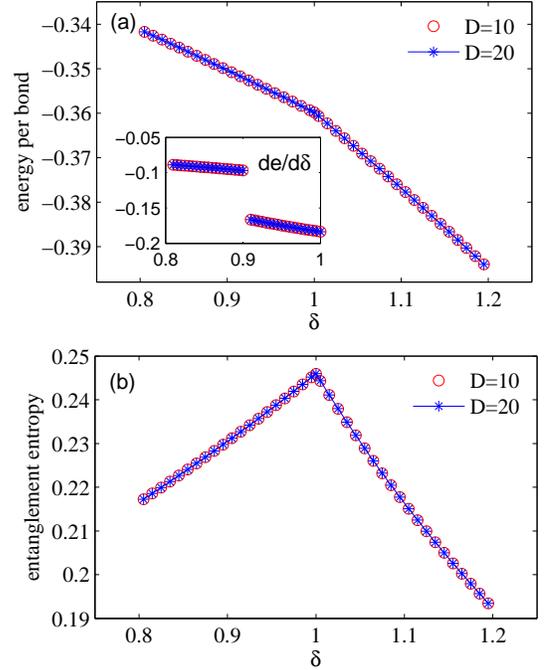}
\caption{(Color online) (a) The ground state energy and (b) the bipartite entanglement entropy as a function of $\delta$ for the XXZ model. The appearance of the cusp in the energy,\Y{as well as the discontinuity in the first-order energy derivative [inset in (a)]}, suggests that this is a first-order phase transition point.}
\label{fig-1st-QPT-XXZ}
\end{figure}

\begin{figure}[tbp]
\includegraphics[angle=0, width=0.9\linewidth]{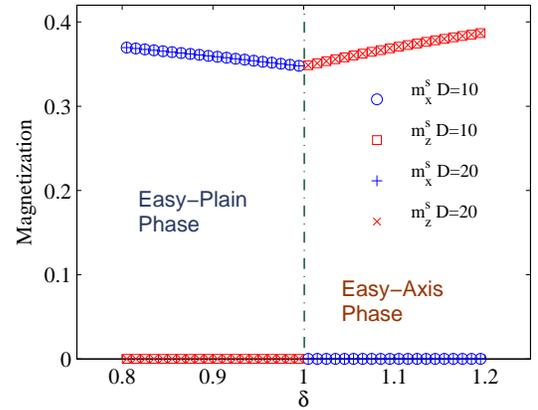}
\caption{(Color online) The staggered magnetization along the z-axis ($m_z^s$) and $x$-axis ($m_x^s$). The spin orientations flip suddenly at the transition point $\delta=1$. For $\delta < 1$, the spins are ordered within the $xy$-plain, while for $\delta > 1$, the ordering is along the $z$-axis.}
\label{fig-magn-XXZ}
\end{figure}

\section{Anisotropic Heisenberg model}

The anistropic Heisenberg model, i.e. the XXZ model, is defined by the Hamiltonian
\begin{equation}
H_{\rm{XXZ}} = \sum_{<i,j>} (S_i^x S_j^x + S_i^y S_j^y + \delta S_i^z S_j^z),
\end{equation}
where $\delta$ is the anisotropy parameter.

The above model has been intensively studied on the honeycomb and
square lattices by different numerical methods, which include
exact diagonalization,\cite{Lin} quantum Monte
Carlo,\cite{Lin, Barnes} coupled cluster
methods,\cite{Bishop1,Bishop2} and tensor network
algorithms.\cite{Chen, Bauer} It is found that the system possesses
magnetic long-range orders for all values of $\delta$.  The
antiferromagnetic ordering vector points within the easy $xy$-plane
for $\delta < 1$ or along the $z$-axis for $\delta > 1$. There is a
first-order phase transition at $\delta = 1$, the Heisenberg
point.\cite{Hajj}

This model was also studied on the Bethe lattice (more precisely, on the Cayley tree lattice) 
by DMRG,\cite{Otsuka, Friedman, Kumar, Changlani}.
It was found that there exists a long-range magnetic order
at the isotropic point $\delta=1$. It was also suggested that a quantum phase
transition occurs at this point. However, the properties of this
transition and the phases on the two sides of the critical point have
not been clarified.

Figure \ref{fig-1st-QPT-XXZ} shows the $\delta$-dependence of the
ground state energy per bond and the entanglement entropy for
the XXZ model.  A clear first order quantum phase transition is
observed at $\delta=1$.  The energy per bond shows a change of slope at $\delta=1$ (\Y{the first-order 
energy derivative is shown in the inset}), which suggests
that there is an energy level crossing. The entanglement entropy
varies continuously across the transition point, but exhibits a
cusp.

Figure \ref{fig-magn-XXZ} shows the staggered magnetization $m_x^s$
and $m_z^s$ around the critical point. The ground state is found to
possess in-plane antiferromagnetic order with a finite
$m^s_x$ for $\delta < 1$, and $z$-axis
antiferromagnetic order with a finite $m^s_z$ for $\delta > 1$. At the transition point,
the two order parameters change suddenly, displaying a spin flip
transition. This result verifies the conjecture made by
Otsuka. \cite{Otsuka}

At the transition point, we find that the ground state energy per bond
has the value $e_b = -0.359817(3)$ and the spontaneous
magnetization has the value $m_s = 0.34736(1)$ for $D=40$. The
errors in the parentheses are estimated by comparing the results for
different bond dimension $D$ and different Trotter slices $\tau$. Our
results agree well with the DMRG data published in
Ref. [\onlinecite{Kumar}], where the local magnetization is found to
be $m=0.347$ on the central lattice site and the bond energy between
the central spin and a spin on the first layer is $e=-0.359$. This satisfactory agreement
suggests that by calculating the Bethe lattice, we can reproduce the
results of local properties in the very center of a large Cayley tree.

\begin{figure}[tbp]
\includegraphics[angle=0, width=1.0\linewidth]{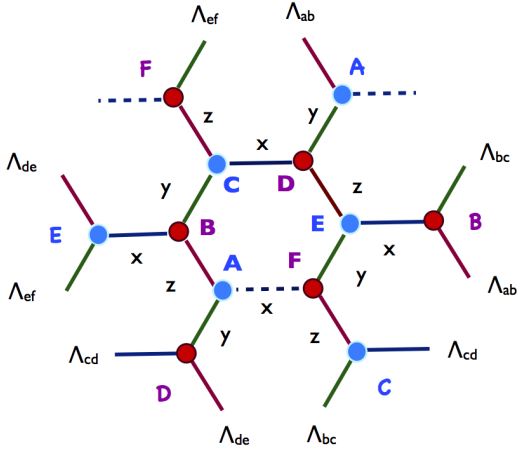}
\caption{(Color online) The one-ring cluster with 12 sites. The diagonal matrices $\Lambda$ are defined on the dangling bonds. The dashed lines represent removed bonds (the physical couplings on the dashed line still exist).}
\label{fig-one-ring}
\end{figure}

\begin{figure}[tbp]
\includegraphics[angle=0, width=1.0\linewidth]{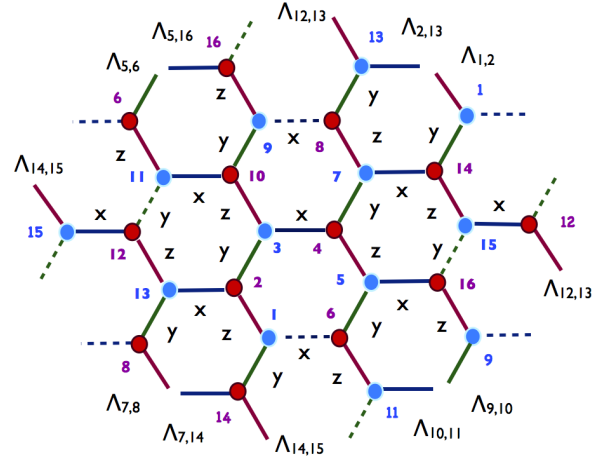}
\caption{(Color online) The 4-ring cluster with 26 sites. The inequivalent lattice sites are numbered from 1 to 16. $\Lambda_{i,j}$ labels the diagonal matrix on the bond linking sites $i$ and $j$.}
\label{fig-four-ring}
\end{figure}

\begin{figure}[tbp]
\includegraphics[angle=0, width=1.0\linewidth]{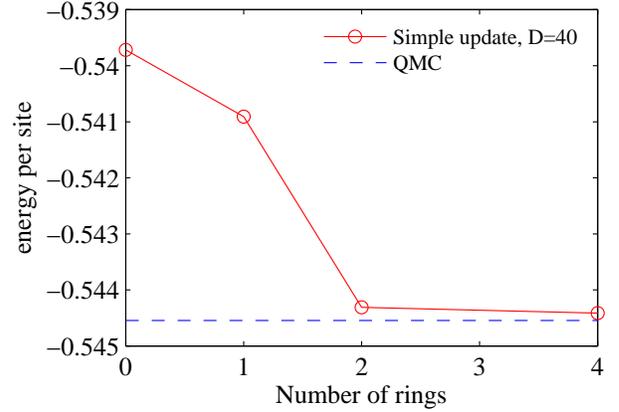}
\caption{(Color online) The ground state energy per site for the Heisenberg model on the honeycomb lattice. The results are evaluated in the central area of the clusters. The dashed line is the recent quantum Monte Carlo result.\cite{Low} The number of hexagonal rings is used to label the cluster size.}
\label{fig-eng-clust}
\end{figure}

\section{Cluster update scheme}

In the previous sections, the simple update has been applied to study the quantum spin models on the Bethe lattice, leading to very accurate results. What is more, in terms of the Bethe approximation, these results can also be regarded as approximations for the corresponding 2D lattice models. Actually, the simple update scheme has already been used to study regular 2D lattices, such as the honeycomb or square lattices. Combined with the TRG/SRG techniques, it can achieve rather accurate results.\cite{Xiang,Xie,Zhao} Nevertheless, in this section, we would provide a different way of using the simple update to calculate 2D lattices. Inspired by the generalization of the Bethe approximation to larger clusters in classical statistical mechanics,\cite{Plischke} we apply simple update to various tree tensor clusters. In this way, the advantage (efficiency) in treating a tree tensor network, namely the fact that it can be readily canonicalized, is utilized to improve the calculation accuracy on a regular 2D lattice.

To start, as a first-order approximation, let
us compare the results on the $q=3$ Bethe lattice (which has no loops)
with those on the 2D honeycomb lattice (whose coordination number is also $q=3$, and it does have loops).
Our result for the ground state energy of the Heisenberg
model on the $q=3$ Bethe lattice is $e_b = -0.359817(3)$, while the
corresponding energy on the honeycomb lattice obtained by the recent
quantum Monte Carlo calculation is
$e_{\rm{QMC}}=-0.36303(14)$. \cite{Low} The relative difference
between these two energies is less than $0.9\%$.  However, the
spontaneous magnetization for the ground state of the Heisenberg model
on the Bethe lattice, namely $m_s = 0.34736(1)$, is much larger
than the corresponding value on the honeycomb lattice, which is about
0.27 as obtained by the quantum Monte Carlo.\cite{Low} Notice, some
other results for the magnetization on the honeycomb
or square lattice obtained with the tensor network algorithms are also
found to be higher than the Monte Carlo ones.\cite{Xiang,Bauer}

As a next step, our approximate treatment of the 2D honeycomb
lattice can be improved by using tensor networks that include rings.
In Fig. \ref{fig-one-ring}, the cluster with one hexagonal ring is
shown, some geometric bonds are removed (dashed lines in
Fig. \ref{fig-one-ring}) to form a tree tensor cluster. Note although
the tensor network does not have geometric
bonds on the dashed lines, in the Hamiltonian the couplings along
these bonds nevertheless exist. Therefore, the projections by
imaginary time evolution should be executed also on the dashed lines.
This cannot be done directly as on usual bonds, but can be accomplished as follows with the help of the swap gates. 
The swap gates are used to exchange the physical indices of two tensors,
which proceeds similarly as the projection scheme illustrated in Fig. \ref{fig-CTN},
with the minor revision that the imaginary tim evolving operator $U(\tau)$
is now replaced with a swap operator $U_s$,
that conducts $U_s | m_i, m_j \rangle = | m_j, m_i \rangle$.

In Fig. \ref{fig-one-ring}, take the dashed bond between site A and F as
an example, swap gates moves the physical index on site A in the order
A $\to$ B $\to$ C, and the physical index on site F as F $\to$ E $\to$
D. After that, the two spins are linked by the solid bond between C and D,
then we can take the projection and update processes as on an usual
bond. After that, we have to move the two spin indices back to their
original positions by reversed swap operations, which accomplishes the
special projection step on a dashed bond. Through iterative and
self-consistent projection processes on the solid and the dashed bonds
of the tree cluster, an approximation for 2D lattices can be obtained.
Compared with the simple Bethe lattice, this tree tensor cluster
approach can provide better approximation for 2D. On the other hand,
it can also be regarded as an ideal method for evaluating the ``super Bethe
lattice'', of which each ``single site'' is now placed with a
hexagonal ring, instead of a single site, and the coordinate number $q=6$.

Beyond the one-ring cluster, more rings can be included to further
enlarge the cluster. As an example, Fig. \ref{fig-four-ring} shows a
cluster with 4 hexagonal rings. The accuracy of energy versus
different cluster size (labelled by the number of rings included) are
shown in Fig. \ref{fig-eng-clust}, which verifies that the accuracy
could be improved consistently with enhancing the cluster size. To
obtain better approximation for true 2D lattices, the local
observables are detected in the center area of the cluster. In
practice, for the 4-ring tree tensor cluster in
Fig. \ref{fig-four-ring}, the results are obtained by averaging over sites 3 and 4.
We find an energy per site of $e \simeq
-0.54441$ (bond energy $e_b \simeq -0.36294$) and a local
magnetization of $m = [e(h_s)-e(h_s=0)]/h_s \simeq 0.3147$ (with a staggered magnetic field $h_s=0.01$).
Hence, the inclusion of rings clearly improves the agreement with
QMC data. For the transverse Ising model, through the 4-ring cluster calculations, the phase transition
point is estimated as $h_c \simeq 1.1$, which is also more accurate
than the simple Bethe approximation. More numerical results with
larger clusters and further details of this cluster Bethe
approximation will be published separately.

\section{Conclusion}

In summary, the simple update scheme is employed to study two spin
models on the Bethe lattice, i.e., the transverse Ising and the Heisenberg
XXZ model. For the Ising model, it is found that the correlation
length, as well as the entanglement entropy, does not diverge at the
second-order transition point. Through a scale transformation, we
have given an intuitive explanation of this peculiar ``critical''
phenomenon. Moreover, by studying the magnetic susceptibility,
we show that the correlation length is upper bounded.
For the Heisenberg XXZ model, the existence of a
first-order phase transition at the isotropic point is clearly
verified, and the two different magnetic
ordered phases are identified as the easy-plain and easy-axis phases,
respectively. Furthermore, in terms of the Bethe approximation, we
obtain accurate and scalable approximations for the 2D lattice
models by applying the simple update to tree tensor clusters.

\section{Acknowledgement}

The authors would like to thank Tomotoshi Nishino and Zhong-Chao Wei for stimulating discussions.
WL is also indebted to Gang Su, Cheng Guo, Guang-Hua Liu, Ming-Pu Qin, Li-Ping Yang,
and Hui-Hai Zhao for helpful discussions. TX was supported by the National Natural
Science Foundation of China (Grants No. 10934008 and No. 10874215) and
the MOST 973 Project (Grant No.  2011CB309703). WL was supported
by the DFG through SFB-TR12.

\end{document}